\documentclass[letterpaper]{article} 
\usepackage{arXiv}  
\usepackage{times}  
\usepackage{helvet}  
\usepackage{courier}  
\usepackage[hyphens]{url}  
\usepackage{graphicx} 
\usepackage{natbib}  
\usepackage{caption} 
\usepackage{algorithm}
\usepackage{algorithmic}
\usepackage{xcolor}
\usepackage{makecell}
\usepackage{multicol}
\usepackage{booktabs}
\usepackage{multirow}
\usepackage{xspace}
\usepackage{amsmath}
\usepackage{textcomp}
\usepackage{verbatimbox}

\newcommand{\LLMs}[0]{\textsc{LLMs}\xspace}
\newcommand{\LLM}[0]{\textsc{LLM}\xspace}
\newcommand{\LLMCs}[0]{\textsc{LLMCs}\xspace}

\newcommand{\Granite}[0]{\textsc{Gra\-ni\-te\-Co\-de}\xspace}

\newcommand{\Qwen}[0]{\textsc{Qwen\-2.5-Co\-der}\xspace}

\newcommand{\Llama}[0]{\textsc{Lla\-ma\-3.2}\xspace}

\newcommand{\DeepSeekCoder}[0]{\textsc{Deep\-Seek-Coder-V2}\xspace}

\newcommand{\LiveCodeBench}[0]{\textsc{LiveCodeBench}\xspace}

\newcommand{\Refactory}[0]{\textsc{Refactory}\xspace}

\newcommand{\ESBMCPython}[0]{\textsc{ESBMC-Python}
\xspace}

\newcommand{\cbmc}[0]{\textsc{CBMC}\xspace}
\newcommand{\cython}[0]{\textsc{Cython}\xspace}
\newcommand{\shedskin}[0]{\textsc{shedskin}\xspace}
\newcommand{\cfaults}[0]{\textsc{CFaults}\xspace}
\newcommand{\tool}[0]{\textsc{Py\-Veri\-tas}\xspace}

\usepackage[newfloat,frozencache,cachedir=.]{minted}

\usepackage{tikz}
\usetikzlibrary{arrows,automata,backgrounds,calc,chains,external,fit,graphs,patterns,positioning,shapes,shapes.geometric}

\tikzset{
    >=stealth',
    punkt/.style={
           rectangle,
           rounded corners,
           draw=black, very thick,
           text width=10em,
           minimum height=2em,
           text centered},
    pil/.style={
           ->,
           thick,
           shorten <=2pt,
           shorten >=2pt,},
    ionode/.style={
           rectangle,
           rounded corners,
           text width=12em,
           minimum height=4em,
           text centered},
}

\tikzset{fancy/.style={rectangle,
		rounded corners=1mm,
		ultra thin,
		draw=white,
		top color=white,
		bottom color=black!20,
		draw}}

\definecolor{highClr}{rgb}{1.0, 1.0, 0.0}

\colorlet{edgeClr}{orange!80!black}
\tikzset{sandEdge/.style={
		>=stealth,
		shorten >=1pt,
		thick,
		bend left,
		text=black,
		edgeClr,
	}}

\tikzset{fadedEdge/.style={
		->,
		>=stealth,
		shorten >=1pt,
		thick,
		edgeClr!20,
	}}

\tikzset{weigthLabel/.style={
		text=black,
		sloped,
		midway,
		}}

\tikzset{fadedWeigth/.style={
		text=lightgray!40,
		sloped,
		midway,
		anchor=south,
		}}

\tikzset{blueVertex/.style={
		rectangle,minimum size=6mm,rounded corners=3mm,
		top color=white,bottom color=blue!35!cyan!25!,
		font=\ttfamily,
		text=black,
	}}

\tikzset{blueVertexG/.style={
		rectangle,minimum size=6mm,rounded corners=3mm,
		top color=white,bottom color=blue!35!cyan!25!,
		font=\ttfamily,
		text=black,
		draw=green,
		thick,
	},
	blueVertexY/.style={
		rectangle,minimum size=6mm,rounded corners=3mm,
		top color=white,bottom color=blue!35!cyan!25!,
		font=\ttfamily,
		text=black,
		draw=yellow,
		thick,
	},
	blueVertexO/.style={
		rectangle,minimum size=6mm,rounded corners=3mm,
		top color=white,bottom color=blue!35!cyan!25!,
		font=\ttfamily,
		text=black,
		draw=orange,
		thick,
	}}

\colorlet{noteClr}{lightgray!30!white!50}

\tikzset{noteBckg/.style={
		rounded corners=8pt,fill=noteClr,
	},
	noteStl/.style={
		font=\scriptsize,
		align=center,
		text=black
	}}

\makeatletter

\newcommand*\mysize{%
  \@setfontsize\mysize{8}{8}%
}
\makeatother
\setminted{fontsize=\mysize}

\usepackage{newfloat}
\usepackage{listings}
\DeclareCaptionStyle{ruled}{labelfont=normalfont,labelsep=colon,strut=off} 
\floatstyle{ruled}
\newfloat{listing}{tb}{lst}{}
\floatname{listing}{Listing}
\title{\tool: On Verifying Python via \LLM-Based Transpilation\\ and Bounded Model Checking for C}
\author{
    Pedro Orvalho,
    Marta Kwiatkowska
}
\affiliations{
    Department of Computer Science\\University of Oxford\\Oxford, UK

}

\usepackage{bibentry}

\begin{document}

\maketitle

\begin{abstract} 
Python has become the dominant language for general-purpose programming, yet it lacks robust tools for formal verification. In contrast, programmers working in languages such as C benefit from mature model checkers, for example \cbmc, which enable exhaustive symbolic reasoning and fault localisation. The inherent complexity of Python, coupled with the verbosity and low-level nature of existing transpilers (e.g., \cython), have historically limited the applicability of formal verification to Python programs. 

In this paper, we propose \tool, a novel framework that leverages Large Language Models~(\LLMs) for high-level transpilation from Python to C, followed by bounded model checking and MaxSAT-based fault localisation in the generated C code. \tool enables verification and bug localisation for Python code using existing model checking tools for C. 
Our empirical evaluation on two Python benchmarks demonstrates that \LLM-based transpilation can achieve a high degree of accuracy, up to 80--90\% for some \LLMs, enabling effective development environment that supports assertion-based verification and interpretable fault diagnosis for small yet non-trivial Python programs. 

\end{abstract}


\section{Introduction}

Automated software verification through techniques such as model checking~\cite{Clarke97-model-checking,DBLP:conf/dac/ClarkeKY03,cav12-ic3} has long been a central topic in programming languages and software engineering research, with widespread industrial adoption in safety-critical systems~\cite{ress12-model-checking-nucler-engineering,clarke2000-verification-safety-critical}, hardware verification~\cite{ClarkeK03-hardware-verification}, and web-services \cite{DBLP:conf/icse/ChongCKKMSTTT20,isorc05-model-checking-web-services}. 
In particular, bounded model checking~(BMC)~\cite{bounded-model-checking-2003} has proven to be a highly effective method for exhaustively analysing program behaviours, within fixed bounds on variables and computation steps.
Mature tools such as \cbmc~\cite{tacas04-cbmc-ClarkeKL} have demonstrated remarkable success in verifying and analysing C programs through symbolic reasoning and constraint solving.

Despite advances,
a significant gap remains: there is still no robust, scalable model checker for Python, which is in widespread use today. Recently, ESBMC-Python~\cite{esbmc-python} model checker was proposed; 
however, it remains at the prototype stage, limited to verifying simple functions and supporting only a restricted subset of the language.
Although lightweight static analysis and type-checking tools exist, formal verification of general-purpose Python programs remains largely under-developed. This limitation arises from Python’s highly dynamic and expressive nature, which make it particularly challenging to encode Python semantics within a formal verification framework. 
An alternative route via code transpilation, that is, translating source code from one programming language to another, is impractical for Python. 
Existing transpilers,~such~as~\cython~\cite{cython}~and~\shedskin~\cite{shedskin}, generate thousands of lines of low-level C code even for modest Python functions, preventing effective symbolic analysis.

Recently, \emph{Large Language Models} (\LLMs) trained for code (\LLMCs) have demonstrated strong performance across a range of code-related tasks, including program synthesis~\cite{cav24-yixuan-guiding-program-synthesis-llms}, program repair~\cite{aaai23-repair-multilingual-LLMs,oopsla24-PyDex,aaai25-LLM-Driven-CEGIS}, and code transpilation~\cite{aaai24-code-translation-models,ase24-supervised-code-translaiton,tosem24-BatFix}. 
\LLMCs trained on large-scale multilingual code corpora have shown the ability to produce semantically faithful translations between languages such as Python and C, often achieving more concise and high-level translations than traditional transpilers~\cite{aaai24-code-translation-models}.

In this paper, we propose a novel approach that leverages \LLM-based transpilation from Python to C, followed by automated verification of the generated C code using bounded model checking (via \cbmc). As concrete applications of this approach, we demonstrate how, based on the transpiled C code, one can: (1) verify the correctness of Python programs using \cbmc, and (2) localise bugs in Python code by applying formula-based fault localisation techniques to the C translation using \cfaults~\cite{CFaults-FM24}, a MaxSAT-based fault localisation tool for C programs.
We implement this pipeline in a tool called \tool. This tool offers an interim solution to the problem of Python verification by integrating \LLM-based transpilation with mature C verification tools, bridging the gap until native Python model checking becomes feasible.

\begin{table*}[t!]
\begin{minipage}[t!]{1\columnwidth}
\centering
\begin{algorithm}[H]
\caption{Python function \texttt{distributeCandies(n: int, limit: int)}.}
\label{code:py-distribute-candies}
\begin{minted}[escapeinside=||,tabsize=0,obeytabs,linenos,xleftmargin=0pt]{python}
def distributeCandies(n: int, limit: int) -> int:
    limit = min(limit, n)
    ans = 0
    |\textcolor{red}{ans = 0 + 1}|
    for i in range(limit + 1):
        if n - i > limit * 2:
            continue
        ans += min(limit, n-i)-max(0, n-i-limit)+1
    return ans

assert distributeCandies(n = 5, limit = 2) == 3
\end{minted}
\end{algorithm}
\end{minipage}
\hfill
\hfill
\begin{minipage}[t!]{1\columnwidth}
\centering
\begin{algorithm}[H]
\caption{C version of Algorithm~\ref{code:py-distribute-candies} transpiled by \Qwen from Python to C.}
\label{code:c-distribute-candies}
\begin{minted}[escapeinside=||,tabsize=0,obeytabs,linenos,xleftmargin=0pt]{c}
int distributeCandies(int n, int limit) {
    limit = (limit < n) ? limit : n;
    int ans = 0;
    |\textcolor{red}{ans = 0 + 1;}|
    for (int i = 0; i <= limit; i++) {
        if (n - i > limit * 2) {
            continue;
        }
        ans += ((limit < n-i) ? limit : (n-i)) - 
        ((n-i-limit > 0) ?(n-i-limit) : 0) + 1;
    }
    return ans;
}

int main() {
    assert(distributeCandies(5, 2) == 3);
    return 0;
}
\end{minted}
\end{algorithm}
\end{minipage}
\end{table*}

Our experiments focus on \LLMs with up to 32~billion
parameters on two well-studied Python benchmarks: \LiveCodeBench~\cite{livecodebench} and \Refactory~\cite{ase19-refactory}. We evaluate four different models: three trained specifically for coding tasks, \Qwen, \DeepSeekCoder and \Granite; and one general-purpose model, \Llama, for comparison.
Our experiments show that \LLM-based transpilation can produce semantically faithful C code for Python programs, enabling formally verified Python code in over 80\% of cases with some models, notably \Qwen. 
Moreover, by applying MaxSAT-based fault localisation to the transpiled code, \tool can effectively identify faults injected into the Python source code. 

In summary, this paper makes the following contributions.

\begin{itemize}
\item We propose \tool, a novel framework for verification and fault localisation for Python programs by leveraging \LLM-based transpilation to C.
\item \tool combines \LLM-based code transpilation, bounded model checking with \cbmc, and MaxSAT-based fault localisation using \cfaults.    
\item We conduct an empirical evaluation on two Python benchmarks, showing that \tool accurately verifies small yet non-trivial Python programs where native verification tools fall short.
\item \tool can map localised faults in transpiled C code back to the original Python source, providing interpretable diagnostic feedback.
\end{itemize}

\section{Motivation}

Consider the function \texttt{distributeCandies(n: int, limit: int)} presented in Algorithm~\ref{code:py-distribute-candies}, which computes the total number of ways to distribute \texttt{n} candies among three children such that no child receives more than \texttt{limit} candies. This Python function corresponds to program \#463 from \LiveCodeBench and was obtained from \textsc{LeetCode}~\cite{leetcode}, although line~4, \texttt{ans = 0 + 1},  was introduced as an intended fault by duplicating the assignment on line~3 and adding a constant.
Now, suppose we want to verify that this program satisfies~the~assertion~on~line~11.

Attempting to verify Algorithm~\ref{code:py-distribute-candies} using \ESBMCPython~\cite{esbmc-python}, version 7.9, a recently proposed model checker for Python, proves infeasible. This prototype can only handle a small subset of Python and
its functionality is currently limited, supporting only top-level \texttt{for} loops (i.e., loops cannot be inside functions), restricting lists and strings, and lacking support for dictionaries. Consequently, it is currently not able to analyse the structures in this program.
Alternatively, one might consider using a robust Python-to-C transpiler such as \cython~\cite{cython}. However, this approach is also impractical. \cython translates this 9-line Python function into over 6,100 lines of low-level and complex C code. Analysing and verifying such code would be extremely challenging, and tracing fault information back to the original Python source would be nearly impossible, making it of little help to programmers.

This is where we believe \tool can offer a practical solution, particularly given that existing Python model checkers are still in early development and do not yet support the full expressiveness of Python. Our approach is to use \emph{Large Language Models} (\LLMs) to transpile a Python program to C, whether correct or buggy, guided by a natural language description and a formal specification of the intended behaviour. Once in C, we can apply several mature verification and symbolic analysis tools developed for C to support developers by providing verification and diagnostic feedback at the level of Python.

For instance, if we provide Algorithm~\ref{code:py-distribute-candies} along with its natural language description to \Qwen~\cite{Qwen-2024}, an \LLM fine-tuned for coding tasks, and request a C translation using our transpilation prompt (see Appendix~\ref{sec:prompts}), the model produces the C code shown in Algorithm~\ref{code:c-distribute-candies}. We can then apply \cbmc, a bounded model checker for C, and verify that the assertion fails, confirming that the program is indeed incorrect.
Furthermore, given a C version of the original Python program, we can invoke MaxSAT-based fault localisation tools, such as \cfaults~\cite{CFaults-FM24}, to identify the root cause of the failure. 
In this case, \cfaults reports that line~4 of the C program in Algorithm~\ref{code:c-distribute-candies} is faulty. We can then provide this information to the \LLM and ask it to map the identified line back to the corresponding faulty statement in the original Python source. \Qwen's response is as follows:
\begin{verbnobox}[\fontsize{9pt}{9pt}\selectfont]
Here are the corresponding buggy Python 
statements from the original program that 
correspond to the buggy C statements 
provided:
```python
    ans = 0 + 1
```
\end{verbnobox}

Thus, we successfully localised line 4 as the fault in the original Python program in Algorithm~\ref{code:py-distribute-candies}, through \LLM-based transpilation, model checking, and MaxSAT-based fault localisation for C. To the best of our knowledge, MaxSAT-based fault localisation techniques have, until now, only been developed for C~\cite{bugAssist-pldi11,CFaults-FM24}, with no equivalent methodology yet proposed for Python. Therefore, at present, this remains the only viable way to leverage these powerful localisation approaches, which are capable of precisely identifying the minimal set of faulty statements in a program.

\section{Preliminaries}

This section defines key concepts used throughout the paper.

\paragraph{Satisfiability.} The \emph{Boolean Satisfiability (SAT)} problem is the decision problem for propositional logic~\cite{biere2009handbook}. The \emph{Maximum Satisfiability (MaxSAT)} problem is an optimization version of the SAT problem. Given a CNF formula $\phi$, the goal is to find an assignment that satisfies the maximum number of clauses~\cite{maxsat-handbook-sat}. In partial MaxSAT, $\phi$ is divided into hard clauses ($\phi_h$) and soft clauses ($\phi_s$). The goal is to satisfy all clauses in $\phi_h$ while minimising the number of unsatisfied soft clauses.

\paragraph{Programs and Verification.} A \emph{program} is sequential, comprising standard statements such as assignments, conditionals, loops, and function calls, each following conventional Python/C semantics. A program is considered \emph{buggy} if an assertion violation occurs during execution with input~$I$. Otherwise, it is correct for~$I$. When a bug is observed, the corresponding \emph{error trace} is the sequence of statements executed by program~$P$ on input~$I$.
A \emph{trace formula (TF)} is a SAT formula that is satisfiable if and only if there exists an execution of the program that terminates in an \emph{assertion violation} while respecting all assume statements~\cite{DBLP:conf/dac/ClarkeKY03}.

\paragraph{Formula-Based Fault Localisation.} Given a faulty program and a test suite with failing cases, \emph{formula-based fault localisation} (FBFL) methods encode the localisation task into an optimisation problem, aiming to identify a minimal set of faulty statements (diagnoses). Typically, FBFL tools encode a program’s trace formula as a MaxSAT formula, and apply the theory of \emph{model-based diagnosis}~(MBD)~\cite{reiter87,ijcai15-Marques-SilvaJI15,ijcai19-ignatievMWM,CFaults-FM24} to enumerate all diagnoses, each corresponding to a potential bug~location. 
Following this theory, the clauses encoding the trace formula of the buggy program correspond to the \emph{hard clauses} of the MaxSAT formula, while the set of healthy (or relaxation) variables correspond to the \emph{soft clauses}, as the goal is to maximise the number of healthy program statements.


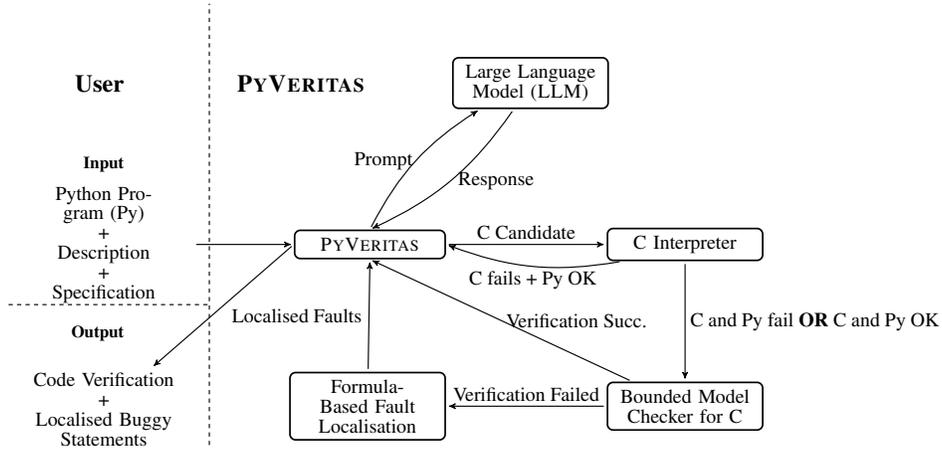
\begin{figure*}[t!]
\centering
\resizebox{1.49\columnwidth}{!}{
\begin{tikzpicture}[node distance=2cm, auto,every node/.style={font=\fontsize{14}{14}\selectfont}]
 \node[punkt] (enum) {\tool};
 \node[punkt, inner sep=5pt, above=3cm of enum, xshift=4cm]
 (llm) {Large Language Model (\LLM)}
 edge[pil, <-, bend left=-15] node[left=-0.05cm]  (l) {Prompt} (enum.north)
 edge[pil, ->, bend left=15] node[right=0.05cm]  (l) {Response} (enum.north);
 \node[punkt, inner sep=5pt,right=4cm of enum]
 (inter) {C Interpreter}
 edge[pil, <-] node[above]  (l) {C Candidate} (enum.east)
 edge[pil, ->, bend left=15] node[below]  (l) {C fails + Py OK} (enum.east);
 \node[punkt, inner sep=5pt,below=3cm of inter]
 (veri) {Bounded Model Checker for C}
 edge[pil, <-] node[right]  (l) {C and Py fail \textbf{OR} C and Py OK} (inter.south)
 edge[pil, ->] node[right] {Verification Succ.} (enum.south);
 \node[punkt, inner sep=5pt,left=4cm of veri]
 (fl) {Formula-Based Fault Localisation}
 edge[pil, <-] node[above] {Verification Failed} (veri.west)
 edge[pil] node[left=0.05cm] {Localised Faults} (enum.south);
\node[ionode,draw=none,left=2.5cm of enum] (specs_node) {Python Program (Py)\\+\\Description\\+\\Specification}
  edge[pil] (enum.west);
\node[ionode,draw=none,below=1.5cm of specs_node] (output) {Code Verification\\+\\Localised Buggy Statements}
  edge[pil, <-] (enum.west);

\draw[dashed, thick] (-4,-5) -- (-4,6);

\draw[dashed, thick] (-9,-1.5) -- (-4,-1.5);

\node[anchor=east] at (0,4) {\textbf{\huge{\tool}}};
\node[anchor=east] at (-6,4) {\textbf{\huge{User}}};
\node[anchor=east] at (-6,2) {\textbf{\Large{Input}}};
\node[anchor=east] at (-6,-2.2) {\textbf{\Large{Output}}};
\end{tikzpicture}
}
\caption{\tool's pipeline for Python verification and bug localisation.}
\label{fig:pyVeritas}
\end{figure*}

\section{\tool}
\label{sec:pyveritas}

In this section, we introduce \tool, a novel framework that enables formal verification and MaxSAT-based fault localisation of Python programs through \LLM-based transpilation to C and model checking for C. 
Figure~\ref{fig:pyVeritas} presents an overview of \tool.

While symbolic tools, for example,  \ESBMCPython~\cite{esbmc-python}, have attempted to support symbolic reasoning over Python, these tools remain in early stages and are limited to simple functions and subsets of the language. On the other hand, bounded model checking (BMC) tools for C, such as \cbmc~\cite{tacas04-cbmc-ClarkeKL}, are mature, efficient, and support fault localisation via MaxSAT-based analysis~\cite{CFaults-FM24,bugAssist-pldi11}.
The goal of \tool is to bridge the gap in formal verification for Python while a robust native model checker for the language is still lacking. \tool operates by translating Python to C programs using Large Language Models (\LLMs), and then leveraging mature C verification infrastructure to perform symbolic analysis and fault localisation. \tool implements a verification pipeline that, given a Python program and its specification, returns either a formal correctness/failure verdict or a set of localised buggy statements mapped back to the original Python source. Please refer to Appendix~\ref{sec:prompts} for all the prompt templates used in the various interactions with the \LLM.

\subsection{Architecture}

\tool takes as input a Python program $\mathcal{P}$, a textual description $\mathcal{D}$ (used to help preserve intent during transpilation), and a set of specifications $\mathcal{S}$, encoded as assertions. Textual descriptions are typically provided to \LLMs to indicate intent, whereas assertions are necessary for formal verification. We remark that this provides redundant information, which may contain inconsistencies, particularly when the code to transpile is faulty.

\paragraph{\LLM-Based Transpilation.} Given $(\mathcal{P}, \mathcal{D}, \mathcal{S})$, \tool constructs a prompt and queries an \LLM to generate a candidate C program $\mathcal{C}$ that aims to preserve the semantics of $\mathcal{P}$ and satisfy the specifications~in~$\mathcal{S}$.





\paragraph{C Interpreter.} Since \LLM-generated code can be imprecise or syntactically/semantically incorrect, \tool executes both $\mathcal{P}$ and $\mathcal{C}$ on the same test inputs and assertions (i.e., $\mathcal{S}$) using a Python interpreter and a C interpreter, respectively. If $\mathcal{C}$ fails to satisfy the assertions in $\mathcal{S}$ 
that $\mathcal{P}$ satisfies, or if $\mathcal{C}$ fails to compile, the candidate  is discarded and a new one is requested from~the~\LLM, up to five failed attempts, indicating the model is likely repeating the same incorrect output; or until the time limit is reached.

\paragraph{Verification via \cbmc.} If the C candidate passes the interpreter, \tool verifies it using a bounded model checker for C, \cbmc~\cite{tacas04-cbmc-ClarkeKL}. If verification succeeds, \tool concludes that the original Python program satisfies the specification $\mathcal{S}$.

\paragraph{MaxSAT-Based Fault Localisation.} If verification fails, the candidate $\mathcal{C}$ is analysed using \cfaults~\cite{CFaults-FM24}, a formula-based fault localisation tool for C. \cfaults encodes the trace formula of $\mathcal{C}$ as a MaxSAT formula and computes a solution that identifies the minimum set of faulty statements in the C program.

\paragraph{Back-Mapping to Python.} \tool maps the identified fault locations in $\mathcal{C}$ back to their corresponding statements in the original Python program $\mathcal{P}$. This mapping is performed by querying the \LLM with the original code and the localised fault locations. The resulting output is the faulty statements localised in the original~Python~code.

\paragraph{Output.} The output of \tool is twofold: (1) if the C code passes verification with \cbmc, the Python program is declared \emph{verified} with respect to $\mathcal{S}$; and (2) if verification fails, the user is presented with a list of localised faulty statements in the original Python source, inferred via MaxSAT-based analysis on the transpiled C code.

\section{Experiments}

The goal of our experiments was to answer the following research questions (RQs).
\textbf{RQ1.} How accurate and reliable are Large Language Models (\LLMs) at transpiling Python programs into semantically equivalent C code?
\textbf{RQ2.} Can MaxSAT-based fault localisation in C, when applied to \LLM-transpiled Python code, effectively identify faults injected into
Python source?
\textbf{RQ3.} How dependable is \LLM-based transpilation as a method for supporting MaxSAT-based fault
localisation in Python via C-level analysis tools?

\paragraph{\textbf{Experimental Setup.}}
All experiments were run using a NVIDIA L40S graphics card with 48GB of memory on an Intel(R) Xeon(R) Gold 5418Y 48-Core CPU Processor and 251GB RAM, using a time limit of 10 minutes.

\paragraph{\textbf{Evaluation Benchmarks.}} 
To evaluate \tool, we use two widely adopted benchmarks of Python programs: \LiveCodeBench~\cite{livecodebench} and \Refactory~\cite{ase19-refactory}.
\LiveCodeBench contains 479 correct programs submitted to programming contests across competition platforms, such as \texttt{LeetCode}~\cite{leetcode}. 
\Refactory is a benchmark consisting of 1783 buggy and 2447 correct Python programs submitted to five different programming assignments. For our evaluation, we focus on correct programs to support our experiment: transpiling the original Python programs to C and formally verifying their correctness using \cbmc. Specifically, we randomly selected twenty-five correct programs from each assignment, resulting in a total of 125 Python programs.

To assess the effectiveness of localising bugs in Python programs using MaxSAT-based fault localisation for C through \cfaults~\cite{CFaults-FM24}, we injected
faults into these Python benchmarks. 
To obtain precise ground-truth information about the location and nature of the bugs in each program, we applied \emph{semantics-altering mutations}, i.e., syntactically valid code transformations that deliberately change program behaviour. Specifically, we used two 
mutations that are easy to inject automatically:  
(1) \emph{Wrong binary operator (WBO)}, where a comparison operator (e.g., \texttt{==}, \texttt{<}, \texttt{>}) is replaced with a different one, thereby flipping the logic of a conditional expression; and 
(2) \emph{Assignment duplication with constant (ADC)}, in which a random assignment is duplicated and a constant is added to the right-hand side of the duplicated statement, introducing an unintended state change.
We ensure that the mutated Python programs semantically differ from the originals by verifying that they fail at least one assertion from the test suite. For each mutation type, we generate a separate transformed version of the benchmark, producing up to one mutated variant per program, with at most one mutation applied.
These code mutations have been previously employed in prior work~\cite{neurips21-bug-detection-repair,fse22-multIPAs,naacl25-SeqCoBench} to augment datasets with buggy programs, enabling the simulation of realistic faults. Although these mutations are simple, they enable a systematic evaluation of whether \tool can accurately identify fault locations via MaxSAT-based localisation.

\paragraph{Large Language Models (\LLMs).}
In our evaluation, we exclusively used open-access \LLMs available on Hugging Face~\cite{huggingface} with at most 32~billions parameters, for two main reasons. First, closed-access models, for example,  \textsc{ChatGPT} and \textsc{DeepSeek}, are cost-prohibitive and raise concerns regarding data privacy. Second, large-scale models (e.g., those with 70B parameters) demand significant computational resources, rendering them unsuitable for local use and out of reach for many developers.
Thus, we evaluated four different \LLMs using the iterative querying setup described in Section~\ref{sec:pyveritas}. Three of these models are \LLMCs, i.e., \LLMs fine-tuned specifically for coding tasks: 
Alibaba's \Qwen~\cite{Qwen-2024}~(32B), 
DeepSeek's \DeepSeekCoder~\cite{DeepSeek-Coder-V2-2024}~(16B), and
IBM's \Granite~\cite{granite-LLM-2024}~(8B).
As a sanity check, we used
Meta's \Llama~\cite{Llama3-LLMs-2024}~(3B),
which is a general-purpose \LLM not specifically trained for coding tasks.
To ensure consistency, all models were run with temperature set to 0.

\begin{table}[t!]
\setlength{\tabcolsep}{1mm}
\fontsize{9}{9}\selectfont
\centering
\begin{tabular}{lrr}
\midrule
\multicolumn{1}{l}{\textbf{Language Model}} & \multicolumn{1}{c}{\textbf{\LiveCodeBench}} & \multicolumn{1}{c}{\textbf{\Refactory}} \\ 
\midrule
\Qwen (32B)                   & 83.7\%                                      & 92.0\%                                  \\
\DeepSeekCoder (16B)              & 65.1\%                                   
& 64.8\%                                  \\
\Granite (8B)                   & 55.9\%                                
& 52.0\%                                  \\
\Llama (3B)                          & 43.0\%                               
& 28.0\%  \\
\bottomrule
\end{tabular}
\caption{Verification success rates for each \LLM on both benchmarks. Percentages indicate the proportion of C programs that were successfully verified by \cbmc and judged semantically equivalent to the original Python code.}
\label{tab:verification}
\end{table}

\subsection{Program Verification for Python}

Our first experiment aimed to assess the reliability of \LLMs in transpiling Python to C, taking into account additional indication of intent, and to evaluate how accurately the verification of the transpiled C code reflects the correctness of the original Python program.

\paragraph{Baseline.} We tried to use \ESBMCPython~\cite{esbmc-python}, version 7.9, as a baseline for our experiments. However, it was unable to verify any of the Python programs from \LiveCodeBench or \Refactory due to unsupported Python features. As discussed in the motivation section, \ESBMCPython currently supports only a limited subset of the language and is designed to verify simple constructs in isolation. When a program combines multiple features, such as, auxiliary functions, or control-flow constructs (e.g., \texttt{for} loops), \ESBMCPython is unable to analyse the program, resulting in failed verification or internal errors.

\paragraph{\tool.} To answer \textbf{RQ1}, we used \cbmc to verify the \LLM-transpiled C code against the provided assertions. All interactions between the evaluated \LLMs and each Python program in both benchmarks were manually inspected to determine whether, in cases where verification succeeded, the transpiled code and assertions were semantically equivalent to the original Python program. This inspection was blind to the identity of the \LLMs. Each model response was independently labelled three times by experts with strong programming backgrounds to ensure consistency, and any discrepancies resolved through discussion. Table~\ref{tab:verification} reports the verification success rate of each model on each benchmark.
\Qwen demonstrates the highest reliability, with 83.7\% success on \LiveCodeBench and 92.0\% on \Refactory, indicating that its C translations are frequently both syntactically valid and semantically faithful to the original Python programs. \DeepSeekCoder follows, though with a marked decline in performance, where its verification success falls to around 65\% on both benchmarks. This contrast indicates that \Qwen more consistently produces semantically faithful and verifiable C translations.
\Granite and \Llama exhibit lower verification success across both benchmarks. Notably, \Granite achieves only 52.0\% on \Refactory. Similarly, \Llama underperforms across both benchmarks, likely due to being a general-purpose model. 
Interestingly, we observe that larger models yield more verifiable translations, suggesting a correlation between model capacity and verification~reliability, which warrants further investigation.
Overall, these results indicate that, although \LLMs vary considerably in reliability, models such as \Qwen already provide sufficiently dependable transpilation to support formal verification of non-trivial Python programs.

\begin{table*}[t!]
\setlength{\tabcolsep}{1mm}
\fontsize{9}{9}\selectfont
\centering
\begin{tabular}{lcccc}
\cmidrule(r){2-5}  
\multicolumn{1}{c}{\textbf{}} & \multicolumn{4}{c}{Bug: \textbf{Wrong Binary Operator (WBO)}} \\ \midrule
\textbf{LLMs} & \textbf{\% Correct Bug Localised} & \textbf{\% Other Bugs Localised} & \textbf{\% Transpiled Fixed Code} & \textbf{\% Compilation Errors} \\ \midrule
\Qwen & \multicolumn{1}{r}{36.2\%} & \multicolumn{1}{r}{14.3\%} & \multicolumn{1}{r}{49.5\%} & \multicolumn{1}{r}{0.0\%} \\
\DeepSeekCoder & \multicolumn{1}{r}{31.4\%} & \multicolumn{1}{r}{19.0\%} & \multicolumn{1}{r}{44.8\%} & \multicolumn{1}{r}{4.8\%} \\
\Granite & \multicolumn{1}{r}{52.4\%} & \multicolumn{1}{r}{29.5\%} & \multicolumn{1}{r}{8.6\%} & \multicolumn{1}{r}{9.5\%} \\
\Llama & \multicolumn{1}{r}{30.5\%} & \multicolumn{1}{r}{23.8\%} & \multicolumn{1}{r}{3.8\%} & \multicolumn{1}{r}{41.9\%} \\
\midrule \midrule
 & \multicolumn{4}{c}{Bug: \textbf{Assignment Duplication with Constant (ADC)}} \\
 \midrule
 & \textbf{\% Correct Bug Localised} & \textbf{\% Other Bugs Localised} & \textbf{\% Transpiled Fixed Code} & \textbf{\% Compilation Errors} \\ \midrule
\Qwen & \multicolumn{1}{r}{37.5\%} & \multicolumn{1}{r}{10.0\%} & \multicolumn{1}{r}{52.5\%} & \multicolumn{1}{r}{0.0\%} \\
\DeepSeekCoder & \multicolumn{1}{r}{50.0\%} & \multicolumn{1}{r}{12.5\%} & \multicolumn{1}{r}{32.5\%} & \multicolumn{1}{r}{5.0\%} \\
\Granite & \multicolumn{1}{r}{45.0\%} & \multicolumn{1}{r}{27.5\%} & \multicolumn{1}{r}{17.5\%} & \multicolumn{1}{r}{10.0\%} \\
\Llama &  \multicolumn{1}{r}{27.4\%} & \multicolumn{1}{r}{15.0\%} & \multicolumn{1}{r}{7.5\%} & \multicolumn{1}{r}{49.9\%} \\ \midrule
 
\end{tabular}
\caption{Results of MaxSAT-Based Fault Localisation with \tool on \Refactory using two fault types: WBO and ADC.}
\label{tab:refactory-fault-loc}
\end{table*}

\begin{table*}[t!]
\setlength{\tabcolsep}{1mm}
\fontsize{9}{9}\selectfont
\centering
\begin{tabular}{lcccc}
\cmidrule(r){2-5}  
\multicolumn{1}{c}{\textbf{}} & \multicolumn{4}{c}{Bug: \textbf{Wrong Binary Operator (WBO)}} \\ \midrule
\textbf{LLMs} & \textbf{\% Correct Bug Localised} & \textbf{\% Other Bugs Localised} & \textbf{\% Transpiled Fixed Code} & \textbf{\% Compilation Errors} \\ \midrule
\Qwen & \multicolumn{1}{r}{16.9\%} & \multicolumn{1}{r}{20.2\%} & \multicolumn{1}{r}{62.6\%} & \multicolumn{1}{r}{0.3\%} \\
\DeepSeekCoder & \multicolumn{1}{r}{16.1\%} & \multicolumn{1}{r}{36.8\%} & \multicolumn{1}{r}{42.9\%} & \multicolumn{1}{r}{4.2\%} \\
\Granite & \multicolumn{1}{r}{41.6\%} & \multicolumn{1}{r}{34.6\%} & \multicolumn{1}{r}{15.2\%} & \multicolumn{1}{r}{8.6\%} \\
\Llama & \multicolumn{1}{r}{22.7\%} & \multicolumn{1}{r}{47.9\%} & \multicolumn{1}{r}{17.7\%} & \multicolumn{1}{r}{11.6\%} \\
\midrule \midrule
 & \multicolumn{4}{c}{Bug: \textbf{Assignment Duplication with Constant (ADC)}} \\
 \midrule
 & \textbf{\% Correct Bug Localised} & \textbf{\% Other Bugs Localised} & \textbf{\% Transpiled Fixed Code} & \textbf{\% Compilation Errors} \\ \midrule
\Qwen & \multicolumn{1}{r}{7.6\%} & \multicolumn{1}{r}{11.0\%} & \multicolumn{1}{r}{81.4\%} & \multicolumn{1}{r}{0.0\%} \\
\DeepSeekCoder & \multicolumn{1}{r}{6.7\%} & \multicolumn{1}{r}{21.4\%} & \multicolumn{1}{r}{69.5\%} & \multicolumn{1}{r}{2.4\%} \\
\Granite & \multicolumn{1}{r}{39.5\%} & \multicolumn{1}{r}{11.9\%} & \multicolumn{1}{r}{36.7\%} & \multicolumn{1}{r}{11.9\%} \\
\Llama & \multicolumn{1}{r}{18.6\%} & \multicolumn{1}{r}{39.0\%} & \multicolumn{1}{r}{34.3\%} & \multicolumn{1}{r}{8.1\%} \\ \midrule
 
\end{tabular}
\caption{Results of MaxSAT-Based Fault Localisation with \tool on \LiveCodeBench using WBO and ADC bugs.}
\label{tab:livecodebench-fault-loc}
\end{table*}

\subsection{MaxSAT-Based Fault Localisation for Python}

To assess the effectiveness of \tool in localising faults in Python programs, we applied MaxSAT-based fault localisation to C code produced by \LLM-based transpilation of two benchmarks: \Refactory and \LiveCodeBench. As described earlier, each correct Python program was first mutated using one of two semantics-altering transformations: \emph{Wrong Binary Operator (WBO)} or \emph{Assignment Duplication with Constant (ADC)}. These mutations are designed to deliberately introduce subtle bugs while preserving syntactic correctness. As described in Section~\ref{sec:pyveritas}, each mutated Python program was then passed through the full \tool pipeline: \LLM-based transpilation, verification using \cbmc, and fault localisation with \cfaults. 
We note that providing redundant information, such as natural language descriptions and explicit assertions, can strongly signal the intended behaviour of a program, guiding the \LLM towards more semantically aligned translations~\cite{verina-arXiv25}.

Tables~\ref{tab:refactory-fault-loc}~and~\ref{tab:livecodebench-fault-loc} present the results for \Refactory and \LiveCodeBench, respectively. For each \LLM and bug type, we report: the percentage of cases where \cfaults successfully identified the injected fault (\%{\it Correct Bug Localised}), the percentage of cases where unrelated valid fault locations were reported (\%{\it Other Bugs Localised}), the percentage of instances in which the \LLM inadvertently corrected the bug during transpilation (\%{\it Transpiled Fixed Code}), and the percentage of transpiled programs that failed to compile~(\%{\it Compilation Errors}).
Since all prompts include both a natural language description and the ground-truth specification, reasoning-oriented models fine-tuned for code often use this information to correct buggy code during transpilation.
This results in the removal of duplicated assignments or the correction of incorrect binary operators, rather than preserving the original faulty logic. 
This observation aligns with findings from prior work~\cite{verina-arXiv25}, which show that providing problem descriptions and specifications improves code generation quality. As this information may contains inconsistencies, our experiments indicate that different LLMs handle these situations differently. We ran an additional experiment (results in Appendix~\ref{sec:no-nl-descriptions}) by excluding the textual descriptions, observing that the rates of bug fixing decline but the underlying behaviour observed previously persists. 
Furthermore, we observe that unrelated localised faults may stem from two main factors. First, incorrect transpilation may produce buggy C code that is not semantically equivalent to the original buggy Python program, causing the fault localiser to highlight different statements. Second, the generated C code may introduce low-level issues, such as memory errors, that are correctly flagged but unrelated to the injected fault.
Please refer to Appendix~\ref{sec:interaction-llms} for examples where fault localisation succeeded, as well as cases where the \LLMs inadvertently fixed the buggy code during transpilation.

For the WBO bug on \Refactory, in Table~\ref{tab:refactory-fault-loc}, \Granite achieved the highest localisation accuracy at 52.4\%, suggesting strong preservation of the faulty semantics. \Qwen and \DeepSeekCoder followed with 36.2\% and 31.4\%, respectively, while \Llama reached 30.5\%. Notably, \Qwen exhibited the highest fix rate (49.5\%), often rewriting the faulty logic into a correct version, whereas \Granite’s low fix rate (8.6\%) indicates stronger fidelity to the original code. \Llama showed the highest compilation error rate (41.9\%), reflecting its general-purpose nature.
For the ADC bug, \DeepSeekCoder performed best on \Refactory (50.0\%), followed by \Granite (45.0\%) and \Qwen (37.5\%). \Llama again had the lowest success rate (27.4\%) and the highest error rate (49.9\%). \Qwen’s 52.5\% fix rate suggests it frequently removed the injected bug. \Granite showed a relatively high rate of unrelated fault reports (27.5\%), due to incorrect code transpilations.
In~Table~\ref{tab:livecodebench-fault-loc}, localisation success rates declined across models on \LiveCodeBench, except for \Granite, which remained strong. For WBO, \Granite again led (41.6\%), while \Qwen and \DeepSeekCoder dropped to 16.9\%~and~16.1\%. \Llama reached 22.7\%, but with 47.9\% unrelated fault reports. \Qwen and \DeepSeekCoder also showed high fix rates (62.6\% and 42.9\%), often repairing the bug during translation. Compilation errors were low overall, except for \Llama~(11.6\%).
For ADC on \LiveCodeBench, the trend persisted. \Granite again led (39.5\%), followed by \Llama (18.6\%). \Qwen and \DeepSeekCoder had the lowest success rates (7.6\% and 6.7\%) and the highest fix rates (81.4\% and 69.5\%), indicating frequent bug removal during translation. \Llama also showed a high rate of unrelated fault~reports~(39.0\%),~due~to~incorrect~transpilations.

Taken together, these results confirm that MaxSAT-based fault localisation on \LLM-transpiled C code is both feasible and effective, answering \textbf{RQ2}. When that transpilation preserves the faulty semantics, \cfaults can successfully identify the root cause of the bug, even though it originated in a Python program. Notably, this holds across different fault types and benchmarks.
In addressing \textbf{RQ3}, we find that the reliability of \LLM-based transpilation for enabling C-level analysis of Python code varies significantly across models. \Qwen achieves the highest verification success, with 92.0\% of programs verified on \Refactory and 83.7\% on \LiveCodeBench, indicating well-structured and semantically coherent output.  In contrast, \Granite yields lower verification rates (52.0\% and 55.9\%), but consistently preserves the faulty semantics, resulting in higher localisation success. 
Thus, we observe that models differ in how they leverage redundant information. Reasoning-oriented models like \Qwen and \DeepSeekCoder tend to generate chains of thought (CoT), integrating cues from natural language descriptions and assertions to guide their output, often repairing faulty code in the process. In contrast, models like \Granite prioritise structural fidelity, preserving the original syntax and semantics (including faults), which is crucial for fault localisation tasks where bug retention is required,
underscoring the importance of model selection when using \LLM-based transpilation to support verification and debugging workflows in \tool.
Moreover, we conducted experiments without providing natural language descriptions to the \LLMs (see Appendix~\ref{sec:no-nl-descriptions}). While localisation accuracy declined slightly, some \LLMs continued to exhibit a tendency to fix the buggy code during~transpilation.

\section{Related Work}

Recent work has explored the use of Large Language Models (\LLMs) for code transpilation and synthesis, particularly across high-level programming languages. Tools such as \textsc{CoTran}~\cite{cotran-ecai24} and \textsc{Miracle}~\cite{ase24-supervised-code-translaiton} leverage neural models for transpiling between languages like Python, Java, and C++, with benchmarks that assess functional correctness. However, these approaches typically focus on transpilation in isolation and do not target the specific Python-to-C direction, nor do they integrate downstream formal verification. Large-scale empirical studies~\cite{llm-based-translation-to-large-scale} reveal that \LLM-generated transpilations often suffer from high error rates (ranging from 50\% to 98\%), and they identify common issues such as semantic hallucinations, inconsistent control flow, and undeclared variables. Nonetheless, these studies are limited to error analysis and do not connect transpilation outputs to formal~guarantees.
Work on verifying C code through deductive or bounded model checking techniques has progressed independently. \textsc{SynVer}~\cite{synver24}, for instance, uses \LLMs to generate C implementations from formal specifications, integrating deductive verification backends. However, it does not address the challenge of transpiling from Python while preserving the semantics or intent of the original code. Moreover, there is currently no demonstrated pipeline that translates Python code into verified C annotated with specifications. Recently, \textsc{BatFix}~\cite{tosem24-BatFix} augmented \LLM-based code transpilers by applying program repair and synthesis to fix faulty translations, producing C++ code from Python or Java that passes all test cases.  \citet{neurips24-verified-code-transpilation} use \LLMs to transpile code into low-level intermediate representations of target languages, while generating formal proofs to ensure semantic equivalence, enabling verified code translation. Finally, 
\cite{pldi25-scalable-large-project-llm-transpilation} used \LLMs to translate large codebases (e.g.\ Go to Rust) by partitioning files, incorporating feature-mapping rules, and validating input-output equivalence to achieve 73\% function-level correctness. 
A separate body of work uses property-based testing and mutation analysis to assess functional equivalence between original and transpiled programs~\cite{ese24-mutation-analysis, aaai24-code-translation-models}. These techniques provide empirical evidence of correctness but typically rely on test generation and lack formal guarantees. 
Finally, several studies have documented the frequent structural and semantic errors introduced by \LLM-based transpilers~\cite{ase24-supervised-code-translaiton, llm-based-translation-to-large-scale}. 
Our work complements this line of research by examining the feasibility of connecting \LLM-based transpilation from Python to C with subsequent formal verification and MaxSAT-based fault localisation, thereby enabling correctness guarantees beyond testing alone. 

\section{Conclusion}

In this paper, we introduced \tool, a framework for verifying Python programs via \LLM-based transpilation to C, followed by assertion-based verification using tools such as \cbmc, and MaxSAT-based fault localisation via~\cfaults. Our experiments surprisingly show that \LLMs can generate semantically faithful C translations for small but non-trivial Python programs, enabling successful verification in over 80\% of cases with some models, notably \Qwen. 
We also demonstrated that MaxSAT-based fault localisation on the transpiled C code can effectively identify injected faults in the original Python source. 
Interestingly, we found that reasoning-oriented models, such as \Qwen and \DeepSeekCoder, often leveraged redundant~cues, such~as natural language descriptions and~assertions, to implicitly repair code during transpilation, while models like \Granite prioritised structural fidelity, making them better suited for MaxSAT-based fault localisation.
These findings suggest that, despite the current limitations of Python model checkers, \tool offers a practical and effective alternative by leveraging mature C verification tools. Furthermore, we demonstrated the feasibility of applying MaxSAT-based fault localisation to Python using simple injected faults as a proof of concept.  As \LLMs continue to improve, we anticipate even broader~applicability to more complex bugs~and~verification~scenarios.

\section*{Acknowledgments}
This project received funding from the ERC under the European Union’s Horizon 2020 research and innovation programme (FUN2MODEL, grant agreement No.~834115) and ELSA: European Lighthouse on Secure and Safe AI project (grant agreement No. 101070617 under UK guarantee).

\bibliography{mybibfile}

\appendix

\section{Prompts}
\label{sec:prompts}

In this section, we present the prompt templates used for all interactions with the Large Language Models (\LLMs).
\begin{itemize}

\item Transpilation prompt:

Given a natural language description (NL\_Description), a Python program (Python code), and one or more Python assertions (assertion), the transpilation prompt asks the model to transpile the Python code and assertion to semantically equivalent C code, as follows: 

\begin{verbnobox}[\fontsize{8pt}{8pt}\selectfont]
Transpile Python to C Code With Assertion: 
You are an exceptionally intelligent coding 
assistant who consistently produces accurate 
and reliable <C code> by transpiling the given 
<Python code> into semantically equivalent 
<C code>. <NL_Description> gives a natural 
language description of the python code. 
Do not forget to  also transpile the given 
Python assertion  into a C assertion!
\end{verbnobox}    

\begin{verbnobox}[\fontsize{8pt}{8pt}\selectfont]
<NL_Description>
{description}
\end{verbnobox}

\begin{verbnobox}[\fontsize{8pt}{8pt}\selectfont]
<Python Code>
```python
{python_code}
{assertion}
```
\end{verbnobox}

\begin{verbnobox}[\fontsize{8pt}{8pt}\selectfont]
<C Code>
```c
\end{verbnobox}

\item Prompt to ask for a different C candidate:

We provide an explanation for why the C translation was incorrect, either indicating that the C code failed to compile (along with the corresponding error message), or that one or more assertions was not successful. This feedback is included in the prompt, and the original transpilation prompt is appended to the end, allowing the model to attempt the translation again.

\begin{verbnobox}[\fontsize{8pt}{8pt}\selectfont]
Your previous code translation was INCORRECT!
Reason: {Reason}
Try again.
{Transpilation Prompt}
\end{verbnobox}    


\item Prompt to map the buggy C statements back to the original Python code:

\begin{verbnobox}[\fontsize{8pt}{8pt}\selectfont]
Map C program statements back to the original python 
program:
We have detected that both the Python and C programs 
are buggy. We have localised the following faulty 
statements in the C program:
{list of faulty C statements}
\end{verbnobox}

\begin{verbnobox}[\fontsize{8pt}{8pt}\selectfont]
Provide us only with the corresponding Python 
statements from the original program that 
correspond to these buggy statements.
```python
\end{verbnobox}

\end{itemize}

\begin{table}[t!]
\setlength{\tabcolsep}{1mm}
\fontsize{9}{9}\selectfont
\centering
\begin{tabular}{lrr}
\midrule
\multicolumn{1}{l}{\textbf{Language Model}} & \multicolumn{1}{c}{\textbf{\LiveCodeBench}} & \multicolumn{1}{c}{\textbf{\Refactory}} \\ 
\midrule
\Qwen (32B)                   & 80.4\%                                      & 87.2\%                                  \\
\DeepSeekCoder (16B)              & 63.5\%                                   
& 61.6\%                                  \\
\Granite (8B)                   & 53.0\%                                
& 49.6\%                                  \\
\Llama (3B)                          & 30.5\%                               
& 22.4\%  \\
\bottomrule
\end{tabular}
\caption{Verification success rates for each \LLM on both benchmarks \textbf{without providing any natural language description}. Percentages indicate the proportion of C programs that were successfully verified by \cbmc and judged semantically equivalent to the original Python code.}
\label{tab:verification-appendix}
\end{table}

\begin{table*}[t!]
\setlength{\tabcolsep}{1mm}
\fontsize{9}{9}\selectfont
\centering
\begin{tabular}{lcccc}
\cmidrule(r){2-5}  
\multicolumn{1}{c}{\textbf{}} & \multicolumn{4}{c}{Bug: \textbf{Wrong Binary Operator (WBO)}} \\ \midrule
\textbf{LLMs} & \textbf{\% Correct Bug Localised} & \textbf{\% Other Bugs Localised} & \textbf{\% Transpiled Fixed Code} & \textbf{\% Compilation Errors} \\ \midrule
\Qwen & \multicolumn{1}{r}{28.6\%} & \multicolumn{1}{r}{32.4\%} & \multicolumn{1}{r}{39.0\%} & \multicolumn{1}{r}{0.0\%} \\
\DeepSeekCoder & \multicolumn{1}{r}{20.0\%} & \multicolumn{1}{r}{49.5\%} & \multicolumn{1}{r}{18.1\%} & \multicolumn{1}{r}{12.4\%} \\
\Granite & \multicolumn{1}{r}{41.0\%} & \multicolumn{1}{r}{32.4\%} & \multicolumn{1}{r}{13.3\%} & \multicolumn{1}{r}{13.3\%} \\
\Llama & \multicolumn{1}{r}{25.2\%} & \multicolumn{1}{r}{26.1\%} & \multicolumn{1}{r}{0.9\%} & \multicolumn{1}{r}{47.8\%} \\
\midrule \midrule
 & \multicolumn{4}{c}{Bug: \textbf{Assignment Duplication with Constant (ADC)}} \\
 \midrule
 & \textbf{\% Correct Bug Localised} & \textbf{\% Other Bugs Localised} & \textbf{\% Transpiled Fixed Code} & \textbf{\% Compilation Errors} \\ \midrule
\Qwen & \multicolumn{1}{r}{22.5\%} & \multicolumn{1}{r}{22.5\%} & \multicolumn{1}{r}{52.5\%} & \multicolumn{1}{r}{2.5\%} \\
\DeepSeekCoder & \multicolumn{1}{r}{25.0\%} & \multicolumn{1}{r}{20.0\%} & \multicolumn{1}{r}{47.5\%} & \multicolumn{1}{r}{7.5\%} \\
\Granite & \multicolumn{1}{r}{42.5\%} & \multicolumn{1}{r}{35.0\%} & \multicolumn{1}{r}{15.0\%} & \multicolumn{1}{r}{7.5\%} \\
\Llama &  \multicolumn{1}{r}{19.9\%} & \multicolumn{1}{r}{15.0\%} & \multicolumn{1}{r}{2.5\%} & \multicolumn{1}{r}{62.3\%} \\ \midrule
 
\end{tabular}
\caption{Results of MaxSAT-Based Fault Localisation with \tool on \Refactory using two fault types: WBO and ADC \textbf{without providing any natural language description}.}
\label{tab:refactory-fault-loc-appendix}
\end{table*}

\begin{table*}[t!]
\setlength{\tabcolsep}{1mm}
\fontsize{9}{9}\selectfont
\centering
\begin{tabular}{lcccc}
\cmidrule(r){2-5}  
\multicolumn{1}{c}{\textbf{}} & \multicolumn{4}{c}{Bug: \textbf{Wrong Binary Operator (WBO)}} \\ \midrule
\textbf{LLMs} & \textbf{\% Correct Bug Localised} & \textbf{\% Other Bugs Localised} & \textbf{\% Transpiled Fixed Code} & \textbf{\% Compilation Errors} \\ \midrule
\Qwen & \multicolumn{1}{r}{29.6\%} & \multicolumn{1}{r}{46.0\%} & \multicolumn{1}{r}{23.5\%} & \multicolumn{1}{r}{0.8\%} \\
\DeepSeekCoder & \multicolumn{1}{r}{18.0\%} & \multicolumn{1}{r}{58.2\%} & \multicolumn{1}{r}{19.4\%} & \multicolumn{1}{r}{4.4\%} \\
\Granite & \multicolumn{1}{r}{42.9\%} & \multicolumn{1}{r}{36.0\%} & \multicolumn{1}{r}{11.4\%} & \multicolumn{1}{r}{9.7\%} \\
\Llama & \multicolumn{1}{r}{27.7\%} & \multicolumn{1}{r}{51.5\%} & \multicolumn{1}{r}{8.3\%} & \multicolumn{1}{r}{12.5\%} \\
\midrule \midrule
 & \multicolumn{4}{c}{Bug: \textbf{Assignment Duplication with Constant (ADC)}} \\
 \midrule
 & \textbf{\% Correct Bug Localised} & \textbf{\% Other Bugs Localised} & \textbf{\% Transpiled Fixed Code} & \textbf{\% Compilation Errors} \\ \midrule
\Qwen & \multicolumn{1}{r}{16.2\%} & \multicolumn{1}{r}{29.5\%} & \multicolumn{1}{r}{53.3\%} & \multicolumn{1}{r}{1.0\%} \\
\DeepSeekCoder & \multicolumn{1}{r}{3.3\%} & \multicolumn{1}{r}{37.6\%} & \multicolumn{1}{r}{54.8\%} & \multicolumn{1}{r}{4.3\%} \\
\Granite & \multicolumn{1}{r}{35.7\%} & \multicolumn{1}{r}{14.3\%} & \multicolumn{1}{r}{37.6\%} & \multicolumn{1}{r}{12.4\%} \\
\Llama & \multicolumn{1}{r}{16.2\%} & \multicolumn{1}{r}{36.2\%} & \multicolumn{1}{r}{30.5\%} & \multicolumn{1}{r}{17.1\%} \\ \midrule
 
\end{tabular}
\caption{Results of MaxSAT-Based Fault Localisation with \tool on \LiveCodeBench using WBO and ADC bugs \textbf{without providing any natural language description}.}
\label{tab:livecodebench-fault-loc-appendix}
\end{table*}

\section{Experiments Without Natural Language Descriptions}
\label{sec:no-nl-descriptions}

To understand the role of natural language descriptions in guiding \LLM-based transpilation, we repeated our experiments without providing any such descriptions to the models. This setup allows us to assess the reliability of \tool when the \LLM must rely solely on the Python code and assertion-based specifications.

As shown in Table~\ref{tab:verification-appendix}, verification success rates declined only modestly for larger models. For instance, \Qwen maintained high success rates at 80.4\% on \LiveCodeBench and 87.2\% on \Refactory (compared to 83.7\% and 92.0\% with descriptions). In contrast, smaller models such as \Llama experienced a more significant drop, with success rates falling to 30.5\% and 22.4\%, respectively. These findings suggest that high-capacity models are still capable of producing semantically faithful C code, even without natural language guidance.

Tables~\ref{tab:refactory-fault-loc-appendix}~and~\ref{tab:livecodebench-fault-loc-appendix} show the results of MaxSAT-based fault localisation under the same conditions. While localisation accuracy declined slightly, some \LLMs continued to exhibit a tendency to fix the buggy code during transpilation. For instance, \Qwen still fixed 52.5\% of \emph{Assignment duplication with constant (ADC)} cases on \Refactory. This suggests that the tendency to repair code is not exclusively driven by natural language cues, but can also emerge from the models' reasoning approach.

Despite the absence of descriptions, \Granite maintained strong localisation performance. For example, it correctly identified the bug in 42.5\% of ADC cases on \Refactory and 39.5\% on \LiveCodeBench, only slightly lower than with descriptions. This further reinforces \Granite{}’s preference for structural fidelity over semantic correction, making it particularly well-suited for fault localisation tasks.

In summary, it appears that natural language descriptions enhance verification and often prompt models to repair faults during transpilation. Removing them reduces this behaviour only marginally, sometimes improving localisation accuracy.

\section{Interactions with Large Language Models~(\LLMs)}
\label{sec:interaction-llms}

In this section, we present representative examples of some interactions with \LLMs.

\begin{itemize}

\item Example of \DeepSeekCoder inadvertently fixing the buggy Python program \#188 from \LiveCodeBench during transpilation to C:

\begin{itemize}
    \item  Natural Language Description: You are given a string \texttt{moves} of length \textit{n} consisting only of characters \texttt{'L'}, \texttt{'R'}, and \texttt{'\_'}. The string represents your movement on a number line starting from the origin 0. In the $i^{th}$ move, you can choose one of the following directions: Move to the left if \texttt{moves[i] = 'L'} or  \texttt{moves[i] = '\_'}; or; move to the right if \texttt{moves[i] = 'R'} or  \texttt{moves[i] = '\_'}. You should return the distance from the origin of the furthest point you can get to after \texttt{n} moves.

\item Python code:

\begin{minipage}[t!]{0.96\columnwidth}
\centering
\begin{algorithm}[H]
\caption{Program \#188 from \LiveCodeBench, with a duplicated assignment with the added constant introduced on line 3.}
\label{code:problem-188}
\begin{minted}[escapeinside=||,tabsize=0,linenos,obeytabs,xleftmargin=0pt]{python}
def furthestDistanceFromOrigin(moves: str)->int:
    left = 0
    |\textcolor{red}{left = 0 + 1}|
    right = 0
    a = 0
    ret = 0
    for c in moves:
        if c == 'L':
            left += 1
        elif c == 'R':
            right += 1
        else:
            a += 1
    ret = max(ret, abs(left - right) + a)
    return ret

assert furthestDistanceFromOrigin(moves='L_RL__R')==3
\end{minted}
\end{algorithm}
\end{minipage}

\item We fed this information to \DeepSeekCoder using our transpilation prompt.

\item \DeepSeekCoder's response:

\begin{verbnobox}[\fontsize{8pt}{8pt}\selectfont]
 ```c
#include <assert.h>
#include <stdio.h>
#include <stdlib.h>

int furthestDistanceFromOrigin(const char* moves){
  int left = 0;
  int right = 0;
  int a = 0;
  int ret = 0;

  for (int i = 0; moves[i] != '\0'; i++) {
      if (moves[i] == 'L') {
        left++;
      } else if (moves[i] == 'R') {
        right++;
      } else {
        a++;
    }
  }

  ret = abs(left - right) + a;
  return ret;
}

int main() {
  const char* moves = "L_RL__R";
  assert(furthestDistanceFromOrigin(moves)==3);
  return 0;
}
```

This C code transpiles the given Python code 
into a C program. It includes the necessary 
C library for assertions (`assert.h`) and 
ensures that the `furthestDistanceFromOrigin` 
function is tested with the provided example 
`moves = "L_RL__R"`, which should return `3`. 
The `main` function includes an assertion to 
check the result.
\end{verbnobox}

\item Thus, \DeepSeekCoder transpiled Algorithm~\ref{code:problem-188} to C, and during this process the model corrected the program's semantics by removing the duplicated assignment with the added constant that we introduced on line 3, \texttt{left = 0 + 1}.

\end{itemize}


\item Example of program \#57 from \LiveCodeBench, where MaxSAT-based fault localisation for C successfully identified a \emph{wrong binary operator (WBO)} in the original Python code:

\begin{itemize}
    \item  Natural Language Description: A truck has two fuel tanks. You are given two integers, \texttt{mainTank} representing the fuel present in the main tank in liters and \texttt{additionalTank} representing the fuel present in the additional tank in liters. The truck has a mileage of \texttt{10} KM per liter. Whenever \texttt{5} liters of fuel get used up in the main tank, if the additional tank has at least \texttt{1} liters of fuel, \texttt{1} liters of fuel will be transferred from the additional tank to the main tank. You should return  the maximum distance which can be traveled. Note: Injection from the additional tank is not continuous. It happens suddenly and immediately for every \texttt{5} liters consumed.

\item Python code:

\begin{minipage}[t!]{0.9\columnwidth}
\centering
\begin{algorithm}[H]
\caption{Program \#57 from \LiveCodeBench, with a wrong binary operator introduced on line 4.}
\label{code:problem-57}
\begin{minted}[escapeinside=||,tabsize=0,linenos,obeytabs,xleftmargin=0pt]{python}
def distanceTraveled(a: int, b: int) -> int:
    res = 0
    while a:
        if |\textcolor{red}{a < 5}|:
            res += 5 * 10
            a -= 5
            if b:
                b -= 1
                a += 1
        else:
            res += a * 10
            a = 0
    return res

assert distanceTraveled(a = 1, b = 2) == 10
\end{minted}
\end{algorithm}
\end{minipage}

\item We fed this information to \Granite using our transpilation prompt.

\item \Granite's response:

\begin{verbnobox}[\fontsize{8pt}{8pt}\selectfont]
Here is the C code that transpiles the given 
Python code:

```c
#include <assert.h>

int distanceTraveled(int mainTank, 
int additionalTank){
    int res = 0;
    while (mainTank) {
        if (mainTank < 5) {
            res += 5 * 10;
            mainTank -= 5;
            if (additionalTank) {
                additionalTank -= 1;
                mainTank += 1;
            }
        } else {
            res += mainTank * 10;
            mainTank = 0;
        }
    }
    return res;
}

int main() {
    assert(distanceTraveled(1, 2) == 10);
    return 0;
}
```

In this C code, we have defined a function 
`distanceTraveled` that takes two integer 
arguments `mainTank` and `additionalTank` 
and returns an integer value. 

The function uses a `while` loop to iterate 
until `mainTank` becomes zero.

Inside the loop, we have used an `if` 
statement to check if `mainTank` is less 
than 5. 
If it is, we add 50 to the `res` variable 
and subtract 5 from `mainTank`. 

If `additionalTank` is not zero, we 
subtract 1 from `additionalTank` and 
add 1 to `mainTank`.

If `mainTank` is not less than 5, we add
the current value of `mainTank` multiplied 
by 10 to the `res` variable and 
set `mainTank` to zero.

Finally, we have defined a `main` function
that uses the `assert` macro from the 
`assert.h` library to check if the 
`distanceTraveled` function returns the 
expected value for the given input.
\end{verbnobox}

\item \Granite transpiled Algorithm~\ref{code:problem-57} to C (note that \Granite actually took the variable names from the description despite renaming the variables during this process); it preserved the program’s semantics by retaining the buggy binary operator on line 4 of Python code.

\item We then invoke \cfaults to perform MaxSAT-based fault localisation on the C program, which identifies line 6 as faulty.

\item This information is then provided to \Granite using our mapping prompt, in order to trace the buggy C statement back to its corresponding line in the original Python program.

\item \Granite's response:

\begin{verbnobox}[\fontsize{9pt}{9pt}\selectfont]
Here are the corresponding Python statements 
from the original program that correspond 
to these buggy statements:
```python
        if a < 5:
```
\end{verbnobox}

\item Thus, \tool successfully localised line 4 of the original Python program using \cfaults.

\end{itemize}

\vspace{1.5cm}
\item On the other hand, \Qwen, when transpiling the same program (\#57 from \LiveCodeBench), fixed the binary operator, thereby removing the injected fault.

\newpage

\item Example of program \#76 from \LiveCodeBench, where MaxSAT-based fault localisation for C successfully identified a \emph{wrong binary operator (WBO)} in the original Python code using \Qwen:

\begin{itemize}
    \item  Natural Language Description: You are given two integers \texttt{num1} and \texttt{num2}. In one operation, you can choose integer \texttt{i} in the range $[0, 60]$ and subtract \texttt{2i + num2} from \texttt{num1}. Return the integer denoting the minimum number of operations needed to make \texttt{num1} equal to $0$. If it is impossible to make \texttt{num1} equal to $0$, return $-1$.

\item Python code:

\begin{minipage}[t!]{0.9\columnwidth}
\centering
\begin{algorithm}[H]
\caption{Program \#76 from \LiveCodeBench, with a wrong binary operator introduced on line 4.}
\label{code:problem-76}
\begin{minted}[escapeinside=||,tabsize=0,linenos,obeytabs,xleftmargin=0pt]{python}
def makeTheIntegerZero(num1:int, num2:int)->int:
    a = 0
    x, y = (num1, num2)
    while |\textcolor{red}{x < y}|:
        x -= y
        a += 1
        b = x
        c = 0
        cc = 0
        z = 1
        while b:
            c += b % 2 == 1
            cc += (b % 2 == 1) * z
            z *= 2
            b //= 2
        if c <= a and cc >= a:
            return a
    return -1
assert makeTheIntegerZero(num1=5, num2=7)==-1
\end{minted}
\end{algorithm}
\end{minipage}

\item We fed this information to \Qwen using our transpilation prompt.

\item \Qwen's response:

\begin{verbnobox}[\fontsize{8pt}{8pt}\selectfont]
To transpile the given Python code into C, 
we need to carefully translate each part of 
the code, including the logic and the assertion. 
Here's the C code equivalent of the provided 
Python code:

```c
#include <stdio.h>
#include <assert.h>

int makeTheIntegerZero(int num1, int num2) {
    int a = 0;
    int x = num1;
    int y = num2;
    
    while (x < y) {
        x -= y;
        a += 1;
        int b = x;
        int c = 0;
        int cc = 0;
        int z = 1;
        
        while (b) {
            c += (b \% 2 == 1);
            cc += (b \% 2 == 1) * z;
            z *= 2;
            b /= 2;
        }
        
        if (c <= a && cc >= a) {
            return a;
        }
    }
    return -1;
}

int main() {
    assert(makeTheIntegerZero(5, 7) == -1);
    printf("All assertions passed.\n");
    return 0;
}
```

### Explanation:

1. **Function Definition**: The function 
`makeTheIntegerZero` takes two integers 
`num1` and `num2` as input and returns 
an integer.

2. **Variable Initialization**: We initialize 
`a` to 0, and `x` and `y` to `num1` and `num2` 
respectively.

3. **While Loop**: The loop continues as long 
as `x` is less than `y`. Inside the loop:
   - `x` is decremented by `y`.
   - `a` is incremented by 1.
   - Variables `b`, `c`, `cc`, and `z` are 
   initialized. `b` is set to `x`, `c` and `cc` 
   to 0, and `z` to 1.

4. **Inner While Loop**: This loop calculates 
the number of 1s in the binary representation 
of `b` (`c`) and the sum of powers of 2 
corresponding to those 1s (`cc`).
   - `c` counts the number of 1s in the binary 
   representation of `b`.
   - `cc` calculates the sum of powers of 2 
   corresponding to the positions of 1s in `b`.

5. **Condition Check**: If `c` is less than or 
equal to `a` and `cc` is greater than or equal 
to `a`, the function returns `a`.

6. **Return Statement**: If the loop exits 
without finding a valid `a`, the function 
returns `-1`.

7. **Main Function**: The `main` function 
includes an assertion to test the function 
with the example provided. If the assertion
passes, it prints "All assertions passed."

This C code should behave the same way as the 
provided Python code.
\end{verbnobox}

\item \Qwen transpiled Algorithm~\ref{code:problem-76} to C; note that it preserved the program’s semantics by retaining the buggy binary operator on line 4 of Python code.

\item We then invoke \cfaults to perform MaxSAT-based fault localisation on the C program, which identifies line 7, \texttt{while (x < y)}, as faulty.

\item This information is then provided to \Qwen using our mapping prompt, in order to trace the buggy C statement back to its corresponding line in the original Python program.

\item \Qwen's response:

\begin{verbnobox}[\fontsize{8pt}{8pt}\selectfont]
Certainly! Let's map the faulty statement in the 
C program back to the corresponding statement in 
the original Python program.
```python
    while x < y:
```
\end{verbnobox}

\item Thus, \tool successfully localised line 4 of the original Python program using \cfaults.

\end{itemize}

\newpage

\item Finally, we present an interaction with an \LLM, \DeepSeekCoder, where the model not only fixed the original bug, a duplicated assignment, during code transpilation, but also introduced a new semantic error due to incorrect translation. This new fault was subsequently identified by \cfaults. This example corresponds to program \#463 from \LiveCodeBench.

\begin{itemize}
    \item  Natural Language Description: You are given two positive integers \texttt{n} and \texttt{limit}. Return the total number of ways to distribute \texttt{n} candies among $3$ children such that no child gets more than \texttt{limit} candies.

\item Python code:

\begin{minipage}[t!]{0.9\columnwidth}
\centering
\begin{algorithm}[H]
\caption{Program \#463 from \LiveCodeBench, with a duplicated assignment introduced on line 4.}
\label{code:problem-463}
\begin{minted}[escapeinside=||,tabsize=0,linenos,obeytabs,xleftmargin=0pt]{python}
def distributeCandies(n:int, limit:int)->int:
  limit = min(limit, n)
  ans = 0
  |\textcolor{red}{ans = 0 + 1}|
  for i in range(limit + 1):
    if n - i > limit * 2:
      continue
    ans += min(limit, n-i)-max(0, n-i-limit)+1
  return ans

assert distributeCandies(n=5, limit=2) == 3
\end{minted}
\end{algorithm}
\end{minipage}

\item We fed this information to \DeepSeekCoder using our transpilation prompt.

\item \DeepSeekCoder's response:

\begin{verbnobox}[\fontsize{8pt}{8pt}\selectfont]
 ```c
#include <assert.h>
#include <stdio.h>

int distributeCandies(int n, int limit) {
    limit = (limit < n) ? limit : n;
    int ans = 0;
    for (int i = 0; i <= limit; i++) {
        if (n - i > limit * 2) {
            continue;
        }
        ans += (limit < n - i) ? limit : 
            n - i - max(0, n - i - limit) + 1;
    }
    return ans;
}

int main() {
    assert(distributeCandies(5, 2) == 3);
    return 0;
}
```

Explanation:
1. The `distributeCandies` function calculates
the number of ways to distribute `n` candies 
among 3 children such that no child gets more 
than `limit` candies.
2. The `main` function includes the `assert` 
statement to check if the `distributeCandies` 
function returns the expected result for the 
given input `n = 5` and `limit = 2`.
3. The `assert` statement in the `main` 
function ensures that the result of 
`distributeCandies(5, 2)` is equal to 3, 
as per the example provided.
\end{verbnobox}

\item \DeepSeekCoder transpiled Algorithm~\ref{code:problem-463} to C, and during this process the model fixed the injected bug, by removing line 4 (i.e., \texttt{ans = 0 + 1}), although it introduced a new semantic bug on line 10:

\texttt{ans += (limit < n - i) ? limit : n - i - max(0, n - i - limit) + 1}. 

This C statements is buggy since it is not semantically equivalent to the Python statements shown on line 8. A semantically equivalent statement would be:

\texttt{ans += ((limit < n-i) ? limit : (n-i)) - 
        ((n-i-limit > 0) ?(n-i-limit) : 0) + 1;}

\item However, \tool continues and invokes \cfaults to perform MaxSAT-based fault localisation on the C program, which identifies line 10 as faulty.

\item This information is then provided to \DeepSeekCoder using our mapping prompt, in order to trace the buggy C statement back to its corresponding line in the original Python program.

\item \DeepSeekCoder's response:

\begin{verbnobox}[\fontsize{8pt}{8pt}\selectfont]
The buggy statement in the C program corresponds
to the Python statements from the original 
program:
```python
    ans += min(limit, n-i)-max(0, n-i-limit)+1
```
\end{verbnobox}

\item Thus, \tool localised line 8 of the original Python program using \cfaults, due to \DeepSeekCoder's incorrect code transpilation.
\end{itemize}

\end{itemize}

\end{document}